\documentclass[aps,prl,reprint,groupedaddress,showpacs]{revtex4-1}
\usepackage{url}
\usepackage{graphicx}
\usepackage{amsfonts}
\usepackage{amsthm}
\usepackage{amsmath}
\usepackage{amssymb}
\usepackage{appendix}
\usepackage{epstopdf}
\usepackage{natbib}
\usepackage{color}

\begin{document}

\title{Nonaxisymmetric linear instability of cylindrical magnetohydrodynamic Taylor-Couette flow}
\author{Adam Child}
\author{Evy Kersal$\acute{\mathrm{e}}$}
\author{Rainer Hollerbach}
\affiliation{Department of Applied Mathematics, University of Leeds, Leeds, LS2 9JT United Kingdom}

\begin{abstract}
 We consider the nonaxisymmetric modes of instability present in Taylor-Couette flow under the application of helical magnetic fields, mainly for magnetic Prandtl numbers close to the inductionless limit, and conduct a full examination of marginal stability in the resulting parameter space. We allow for the azimuthal magnetic field to be generated by a combination of currents in the inner cylinder and fluid itself, and introduce a parameter governing the relation between the strength of these currents. A set of governing eigenvalue equations for the nonaxisymmetric modes of instability are derived and solved by spectral collocation with Chebyshev polynomials over the relevant parameter space, with the resulting instabilities examined in detail. We find that by altering the azimuthal magnetic field profiles the azimuthal magnetorotational instability, nonaxisymmetric helical magnetorotational instability, and Tayler instability yield interesting dynamics, such as different preferred mode types, and modes with azimuthal wave number $m>1$. Finally, a comparison is given to the recent WKB analysis performed by Kirillov $et$ $al$. [Kirillov, Stefani, and Fukumoto, J. Fluid Mech. 760, 591 (2014)]  and its validity in the linear regime.
\end{abstract}
\maketitle

\section{Introduction}
The dynamics of hydrodynamic Taylor-Couette flow, in which fluid is driven between two coaxially rotating cylinders, is well understood. It is known, by Rayleigh's stability criterion, that fluid is stable as angular momentum increases radially outward. However, electrically conducting fluid can be destabilised through the addition of a weak axial magnetic field \cite{velikhov_1959} and it is this magnetorotational instability (MRI) that is proposed to be the mechanism allowing for outward angular momentum transport in accretion disks \cite{Balbus1991}: a turbulent process that cannot be attributed to laminar viscous effects. Since its discovery the MRI has been the subject of numerous publications, for which we refer to the review by \citet{julien2010magnetorotational}. 

Similarly, if one produces a purely azimuthal field by running a current through the fluid, even without rotation one may excite the current-driven Tayler instability (or Taylor-Vandakurov instability) \cite{vandakurov1972theory,tayler1973adiabatic}. This is fundamentally different to the MRI in that instead of receiving energy from differential rotation, it is driven by the imposed current. The Tayler instability (TI) has numerous astrophysical applications, for example, in the stability of stars, as in Tayler's original investigation, in radiative stellar cores for which there is near solid body rotation, and the Tayler-Spruit dynamo \cite{spruit2002dynamo}, for which it provides the $\alpha$-effect. Further industrial applications include liquid metal batteries \cite{weber2015influence}, though it has recently been shown \cite{herreman2015tayler} that the Tayler instability is too weak to disrupt the electrolyte layer in the batteries.

It is reasonable to have a scenario in which there is differential rotation, as well as an imposed axial field and a current generated azimuthal field in the Taylor-Couette problem. The focus of this paper is then, given that MRI and Tayler instabilities can occur in similar parameter regimes, how would the interaction between them affect the resulting instability?

Experiments involving the MRI and Tayler instabilities are of particular interest, with the first involving the standard MRI (SMRI) suggested independently by \citet{rudiger2001mhd} and \citet{goodman2002}. However, it was clear that any experiment would be particularly difficult to realise, due to the relevant parameters being $\mathrm{Rm}=\Omega_{i}r^{2}_{i}/\eta$ and $\mathrm{S}=B_{0}r_{i}/\eta\sqrt{\mu\rho}$, the magnetic Reynolds number and Lundquist numbers, respectively. As later reiterated by \citet{2004hollerbach}, the difficulty arises due to the requirement that $\mathrm{Rm} \sim O(10)$, which, given that laboratory fluids have $\mathrm{Pm}=\eta/\nu \sim O(10^{-6})$ and $\mathrm{Rm}=\mathrm{Re} \mathrm{Pm}$, would require $\mathrm{Re}\sim O(10^{7})$. At such large rotation rates, the Taylor-Proudman theorem states that any flow would be dominated by the conditions at the endplates. It is not surprising that the SMRI has not yet been achieved \cite{nornberg2010observation}.

As noted by \citet{hollerbach2005}, if an additional azimuthal magnetic field, $B_{\phi}=1/r$, is applied to the fluid a new helical type of magnetorotational instability is found for which Rm is no longer the relevant parameter, allowing instability at experimentally feasible values of $\mathrm{Re}\sim O(10^{3})$. This axisymmetric MRI variant [now known as the helical MRI (HMRI)] has been subsequently observed in the PROMISE experiments with good agreement with  theoretical predictions \cite{stefani2006experimental,stefani2009helical}. It has since been shown by \citet{priede2009absolute} that this observed HMRI is self-sustaining, rather than a transient growth as was suggested previously.

A further nonaxisymmetric MRI variant is produced if one now considers the case in which only an azimuthal magnetic field $B_{\phi} = 1/r$ is present. This azimuthal MRI (AMRI), first discussed theoretically by \citet{ogilvie1996non}, has also been shown to depend on the parameters Re and Ha instead of Rm and S \cite{hollerbach2010nonaxisymmetric} and is thus achievable in laboratory fluids. Indeed, recent work has shown experimental evidence for the AMRI \cite{seilmayer2014experimental}.

Further analysis of the different types of MRI, and in particular the AMRI, was included in a WKB approximation performed by \citet{kirillov2014instabilities}. Previous research had shown that the HMRI was only able to exist for rotation profiles between two restrictive Liu limits \cite{liu2006}, with the possibility of instability at Keplerian rotation profiles excluded \cite{priede2011inviscid} for all inductionless forms of the MRI. \citet{kirillov2013extending} showed that the Liu limits can be bypassed via variation of the azimuthal magnetic field's radial profile, allowing inductionless MRI ($\mathrm{Rm}\rightarrow0$) even under Keplerian rotation. Under further investigation \cite{kirillov2014instabilities}, it was shown that this variation in azimuthal field profile leads to a connecting curve between the two Liu limits.

By allowing for these deviations from the previously defined azimuthal magnetic field profile through generating the field by a combination of currents running through the core and fluid, interesting dynamics may occur. It is then  possible to have both MRI and current driven instabilities, which may occur in overlapping sections of parameter space. \citet{rudiger2010dissipative} have examined this possibility for the case $\mathrm{Pm}=1$. More recently, \cite{priede2015metamorphosis} has fully explored this for the axisymmetric HMRI. 

This paper aims to extend these works to include all possible azimuthal magnetic field profiles and the more experimentally appropriate parameter $\mathrm{Pm}=10^{-6}$ respectively for nonaxisymmetric instability. We derive a set of perturbation equations that incorporate the azimuthal magnetic field configuration in terms of the ratio of current run through the core and fluid, and  discuss the effect this has on onset of instability, both in general and in conjunction with changes to the rotation profile. Furthermore, we comment on the applicability of the WKB theory results of Kirillov {\it{et al.}}, such as the extension of the Liu limits in the linear regime. We show that the choice of magnetic field profile can determine the kind of mode of instability excited, with small changes to the ratio of currents fundamentally changing the flow structure. Finally, we examine the effects that axial field strength has on stability, and show evidence of the appearance of higher azimuthal mode-numbers for axially dominant magnetic fields as suggested by \citet{kirillov2014instabilities}.

\section{Mathematical setting}
We consider cylindrical Taylor-Couette flow with radii $r_i$ and $r_o$ and angular velocity $\Omega_{i}$ and $\Omega_{o}$ for the inner and outer cylinders, respectively. In order to allow any results to be examined experimentally, we take the inner and outer radii to be 4 and 8cm, respectively, to match the geometry of the PROMISE experiment \cite{stefani2006experimental}, giving a relative gap width $r_{i}/r_{o} = 0.5$.

Now, let us include a helical basic state magnetic field with components in the azimuthal and axial directions such that the axial field is generated externally and the azimuthal field by running a combination of currents in the core and the fluid itself. We consider a number of possibilities for the current configuration, including running current in solely the core or fluid and running current in both the core and fluid in the same, or opposite, direction.

The system of equations we derive reduces to a one-dimensional generalised eigenvalue problem, however the inherent difficulty arises from the number of control parameters that govern the setup. We have freedom in choosing the angular velocity of both the inner and outer boundaries; $\Omega_i$ and $\Omega_o$, the relative strength of the axial magnetic field; $B_z$, and the relative strength of the components of the azimuthal field produced by the current in the core and the fluid; $I_{in}$ and $I_{fl}$, respectively.

\subsection{Governing equations}

We consider the basic state of the system, driven by the differential rotation of the two cylinders, as well as the background magnetic field. This gives a basic state velocity  ${\bf U_{0}} = (0,r\Omega{\bf{e}_\phi},0)$, with the standard Taylor-Couette angular velocity profile $\Omega(r)$.
Here the parameter $\mu_{\Omega}$, the rotation ratio, prescribes the steepness of the rotation profile and is given by $\mu_{\Omega}$ = $\Omega_{o}$/$\Omega_{i}$.

The basic state magnetic field ${\bf B_{0}} = (0,B_{\phi},B_{z})$, where $B_{z}$ is the axial magnetic field and $B_{\phi}$ is the azimuthal field produced by running currents in both the inner cylinder $I_{in}$ and the fluid itself $I_{fl}$,

\begin{equation} B_{\phi} = \frac{\mu_{0}}{2\pi}\biggl[\frac{1}{r}\,I_{in}
  + \frac{1}{r}\,\frac{r^2-r_i^2}{r_o^2-r_i^2}\,I_{fl}\biggr].
\label{eq:fieldprofile}
\end{equation}

The nondimensionalizations of length, time and $U$ are all straightforward: Lengths are scaled with $r_i$, basic state velocity with $\Omega_{i}r_{i}$, time on the viscous diffusive timescale $r_{i}^{2}/\nu$, and perturbation velocity with $\eta/r_{i}$. Due to the range of combinations of the azimuthal magnetic fields we wish to examine, scaling the magnetic field is nontrivial: Nondimensionalizations must be valid for zero current in either the fluid or core, as well as equal and opposite currents. The best choice is ultimately to scale both basic state and perturbation magnetic fields with $\bar{B}$, the rms mean of $B_{\phi}$.

Restricting attention to $r_i/r_o=0.5$, this is given by
\begin{equation}
\bar{B}=\frac{\mu_{0}}{2\pi{r_i}}\,I_{in}\,\biggl(\frac{2}{3}\biggr)^{1/2}\,
           \biggl[\ln2\,\Bigl(1-\frac{1}{3}\,\tau\Bigr)^2
                  + \tau + \frac{1}{12}\,\tau^2\biggr]^{1/2},
\end{equation}
where $\tau=I_{fl}/I_{in}$ is the ratio of currents flowing within the fluid and the inner core. The nondimensional equivalent of the field profile (\ref{eq:fieldprofile}) then becomes

\begin{equation}
 B_{\phi} =  \frac{\left(\frac{3}{2}\right)^{1/2}\left(\frac{1}{r}\left(1-\frac{1}{3}\tau\right)+\frac{r\tau}{3}\right)}{\left(\ln2\left(1-\frac{\tau}{3}\right)^2 + \tau + \frac{\tau^{2}}{12}\right)^{1/2}}.
\end{equation}

Note that this profile is well defined for all choices of $\tau$, including $\tau=0$ (corresponding to currents only in the inner core) and $\tau=\pm\infty$ (corresponding to currents only in the fluid). The new nondimensional axial field is simply given the name $\delta$, that is, $B_z=\bar{B}\delta$. 

To summarize, the basic state whose stability is studied in (\ref{maineq1})--(\ref{maineq4}) consists of the usual Couette flow profile $\Omega(r) = c_1 + c_2/r^2$, with $c_1$ and $c_2$ determined by $\mu_{\Omega}$, together with magnetic fields determined by $\tau$ and $\delta$. 

We linearise the governing equations about the basic state, expressing the perturbation flow and magnetic field, $\bf{u}$ and $\bf{b}$, as a toroidal-poloidal decomposition to satisfy the $\nabla\cdot{\bf{u}}=\nabla\cdot{\bf{b}}=0$ conditions,

\begin{align}
\label{udef}
 {\bf u} &= \nabla\times{(e{\bf\hat e}_{r})} + \nabla\times\nabla\times{(f{\bf\hat e}_{r})},\\
 {\bf b} &= \nabla\times{(g{\bf\hat e}_{r})} + \nabla\times\nabla\times{(h{\bf\hat e}_{r})}.
 \label{bdef}
\end{align}

We then expand in terms of normal modes, $e=e(r)\mathrm{exp}(\gamma{t}+im\phi+ikz)$, etc., such that $\gamma$ is the (complex) growth rate and $m$ and $k$ are the azimuthal and axial wave numbers, respectively. Taking the $r$-components of the curl and double curl of the Navier-Stokes equations, as well as the induction equation and its curl, we obtain the following set of eigenvalue equations:

\begin{widetext}
\begin{eqnarray}
\label{maineq1}
{\gamma}(C_{2}e + C_{3}f) + C_{4}e + C_{5}f &=& \mathrm{Re}E_{1} + \mathrm{Re}F_{1} + \mathrm{Ha}^{2}G_{1} + \mathrm{Ha}^{2}H_{1}, \\
{\gamma}(C_{3}e + C_{4}f) + C_{5}e + C_{6}f &=& \mathrm{Re}E_{2} + \mathrm{Re}F_{2} + \mathrm{Ha}^{2}G_{2} + \mathrm{Ha}^{2}H_{2}, \\
{\mathrm{Pm}\gamma}(C_{1}g + C_{2}h) + C_{3}e + C_{4}f &=& E_{3} + F_{3} + \mathrm{Rm}G_{3} + \mathrm{Rm}H_{3}, \\
{\mathrm{Pm}\gamma}(C_{2}g + C_{3}h) + C_{4}e + C_{5}f &=& E_{4} + F_{4} + \mathrm{Rm}G_{4} + \mathrm{Rm}H_{4}, 
\label{maineq4}
\end{eqnarray}
\end{widetext}
where the coefficients are mostly the same as those given by \citet{hollerbach2010nonaxisymmetric}, with the differences being,

\begin{align}
E4 &= i\Delta\left(\frac{mB_{\phi}}{r} + k\delta\right)e,\\
F3 &= i\Delta\left(\frac{mB_{\phi}}{r} + k\delta\right)f,\\
F4 &= ik\left(\Delta{B_{\phi}'} - \frac{B_{\phi}(\Hat{\Delta} - \Delta)}{r} - \frac{2km\delta}{r^{2}}\right)f,\\
G1 &= i\Delta\left(\frac{mB_{\phi}}{r} + k\delta\right)g,\\
G2 &= -ik\Hat{\Delta}\left(\frac{B_{\phi}}{r} + \frac{km\delta}{r^{2}}\right)g,\\
H1 &= -ik\left(\Delta{B_{\phi}'} + \frac{B_{\phi}(\Hat{\Delta} - \Delta)}{r} + \frac{2km\delta}{r}\right)h,\\
H2 &= \frac{im\Delta}{r}\left({B_{\phi}''} + \frac{{B_{\phi}'}}{r} - \frac{B_{\phi}}{r^{2}}\right)h\\
   & + \frac{4imB_{\phi}k^{2}}{r^{3}}h + i\left(\frac{mB_{\phi}}{r} + k\delta\right)C_{4}h.\nonumber
\end{align}

Here, as in \cite{hollerbach2010nonaxisymmetric} we use the notation $\Delta = m^{2}/r^{2} + k^{2}$ and $\hat{\Delta} = 4m^{2}/r^{2} + 2k^{2}$. The nondimensional numbers are the Reynolds number $\mathrm{Re}=\Omega_ir_i^2/\nu$, the Hartmann number $\mathrm{Ha} = \bar{B}r_i/\sqrt{\mu_0\rho\nu\eta}$, and the magnetic Prandtl number $\mathrm{Pm} = \nu/\eta$. Note that each of the new coefficients contains the nondimensional quantities $B_{\phi}$ and $\delta$,  which describe the basic state magnetic field.

The boundary conditions associated with (\ref{maineq1})--(\ref{maineq4}) are taken to be no slip for the flow, i.e.,
\begin{equation}
 e=f=\frac{\partial{f}}{\partial{r}} = 0,
\end{equation}
and insulating for the magnetic field, i.e.\ ${\bf{J}} = \nabla\times{\bf{B}} = {\bf{0}}$ for $r<r_{i}$ and $r>r_{o}$. This yields
\begin{equation}
  \Delta{g} - \frac{2km}{r_i^2}h = 0, 
\end{equation}
and
\begin{align}
\frac{m}{r_{i}}g &= 0,\\
\ k\left(h' + \frac{h}{r_i}\right) - \Delta\frac{I_{m}(kr_i)}{I_{m}'(kr_i)}h &= 0.
\end{align}
at the inner boundary. For the outer boundary $r_{o}$, replace $r_{i}$ with $r_{o}$ and $I_{m}$ with $K_{m}$ respectively. In Fig. 1 we also use perfectly conducting boundary conditions, such that $B_r = \partial{B_{\phi}}/\partial{r} + B_{\phi}/r = 0$, giving
\begin{align}
 h &= 0,\\
 ikg' +\frac{ikg}{r} + \frac{imh''}{r} - \frac{imh}{r^2} &= 0.
\end{align}

We solve (\ref{maineq1})--(\ref{maineq4}) by taking radial expansions of $e, f, g$ and $h$ in terms of Chebyshev polynomials, of degree up to $N=30-80$ depending on the parameter regime, before collocating at $N$ Chebyshev-Gauss-Lobatto nodes. After including the boundary conditions, this gives a $(4N+10)\times(4N+10)$ matrix eigenvalue problem with eigenvalues $\gamma$. We then optimise to obtain the most unstable axial wavenumber $k$. The resulting code has been benchmarked against previous results from both \cite{rudiger2010dissipative,hollerbach2010nonaxisymmetric}, as well as another independently written code.

\section{Results}
\subsection{$\mathrm{Pm}=1$ -- A destabilisation of Chandrasekhar's equipartition solution}

\begin{figure*}[ht]
 \centering
\includegraphics[width=\textwidth]{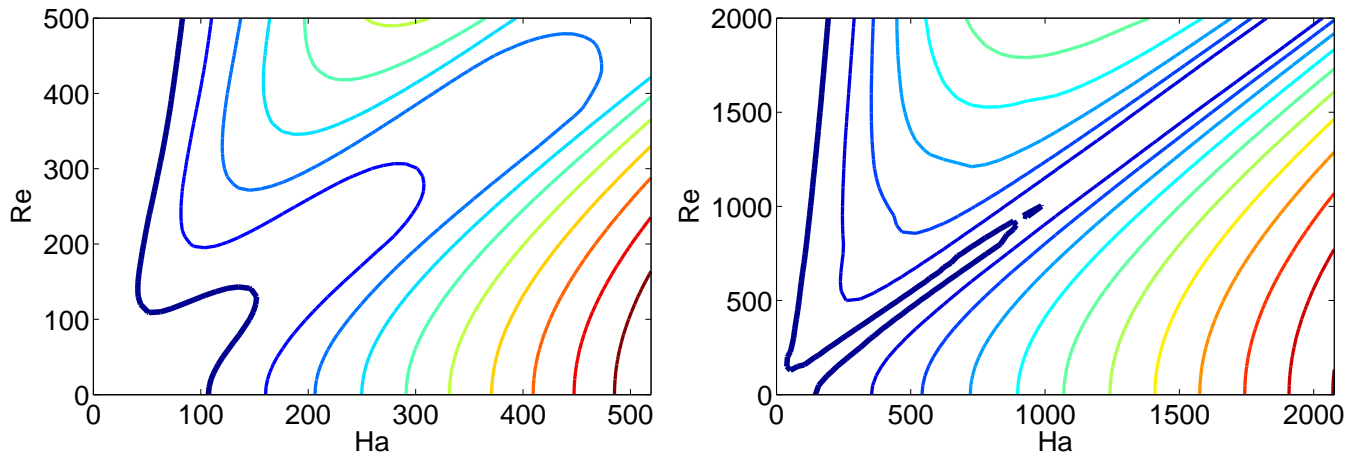}
\caption{(Color online) Weak instability of Chandrasekhar's equipartition between AMRI and TI for Pm=1. Note that in each plot the upper left region of instability is the AMRI and the lower right region is the TI. The contours are taken as the growth rate $\gamma$. Positive growth rates are plotted, with the thickest line indicating the case of marginal stability. Here $\mu_{\Omega}=0.5, \tau=1, m=1$ and $\delta=0$, giving a purely azimuthal field. Left: insulating boundary conditions. Right: perfectly conducting boundary conditions. This compares directly with Fig.\ 7 of \cite{rudiger2010dissipative}, where only the range Ha $\leq 400$, and Re $\leq 500$ is considered.}
\label{fig:Pm1weak}
\end{figure*}

We first consider the case where $\mathrm{Pm}=1$. In their paper, \citet{rudiger2010dissipative} imposed perfectly conducting boundary conditions, and examined a number of different parameter ranges for the AMRI at this value of Pm. One instance used a flat rotation law $\mu_{\Omega}=0.5$, as well as the nearly uniform azimuthal field corresponding to $\tau = 1$. R{\"u}diger $et$ $al$. claimed that for this value of Pm there is a stable separating region between the AMRI and Tayler instabilities where the differential rotation stabilises the Tayler instability. When run over the same Re and Ha range, we recover the results of R{\"u}diger $et$ $al$., however upon considering a larger range for Re and Ha a relatively weak region of instability occurs between the two (Fig.\ \ref{fig:Pm1weak}). This feature is made more apparent when considering insulating boundary conditions, where the weak instability appears for values of Re and Ha inside the range R{\"u}diger $et$ $al$. examined. It has recently been established by \citet{kirillov2014instabilities} that this weak instability is due to the destabilisation of Chandrasekhar's equipartition solution, and it is shown via the analytic WKB solution that the dissipation-induced instability develops into the AMRI as $\mathrm{Rm}\rightarrow0$. This has since been confirmed numerically by \citet{rudiger2014diffusive}. Chandrasekhar's original investigation \cite{chandrasekhar1956stability} showed that when a flow of velocity $U$ is aligned with the magnetic field of Alfv$\acute{\mathrm{e}}$n velocity $U_{Alf}$, $U=U_{Alf}$, the flow is stable. This was shown to be destabilised \cite{mond1987} if $U=\beta{U_{Alf}}$ with some constant $\beta\ne1$, and has since shown to be unstable even for $\beta=1$, albeit not as readily, by considering dissipation. $\beta$, which determines the relative strength of the velocity and Alfv$\acute{\mathrm{e}}$n velocities, can be worked out in our nondimensionalization to be $\beta=\mathrm{Re}\sqrt{\mathrm{Pm}}/\mathrm{Ha}$. 
Here, taking the quasi-galactic rotation rate $\mu_{\Omega}=0.5$ we note that setting $\tau=1$ corresponds to the magnetic field steepness $\mu_{B} = B_{\phi}(r_i)/B_{\phi}(r_o) = 1$ (in R{\"u}diger's notation), resulting in the Chandrasekhar line $2\mu_\Omega=\mu_{B}$. This has radially aligned profiles for $U$ and $U_{Aalf}$, and thus for $\mathrm{Pm}=1$ is the equipartition  $\mathrm{Re}=\mathrm{Ha}$ as seen in Fig. \ref{fig:Pm1weak}. This destabilisation is markedly different in both its location in parameter space and its growth rate when comparing  perfectly conducting and insulating boundaries and adds nicely to the cases of $\mu_{\Omega}=0.25$ and $\mu_{\Omega}=0.35$ examined by \citet{rudiger2014diffusive}, as it shows clearly the comparative weakness of the instability on the equipartition line. 

\subsection*{$\mathrm{Pm}=10^{-6}$ -- Purely azimuthal magnetic fields i.e. $\delta=0$}
\subsection{Changing $\tau$}

We now focus solely on the near-inductionless limit, $\mathrm{Pm}=10^{-6}$, a value typical of liquid metals. We first consider the effect of  changing the parameter $\tau$, corresponding to the ratio of currents running though the fluid and core, when there is no axial field $\delta$ applied. We recall that all values of $\tau$ are valid, with $\tau=0$ corresponding to current free in the fluid, and $\tau=\pm\infty$ corresponding to current solely in the fluid, and focus on the range $-3<\tau<3$. Whereas both \citet{rudiger2010dissipative} and \citet{kirillov2014instabilities} have considered varying magnetic field profiles in this manner, a range such as this has not been examined. In particular, the negative values of $\tau$ such as $\tau<-1$ have not been included when expressing the magnetic field profile in terms of $\mu_{B}$ as in \cite{rudiger2010dissipative}, or the magnetic Rossby number as in \cite{kirillov2014instabilities}.

\begin{figure*}[ht]
 \centering
\includegraphics[width=\textwidth]{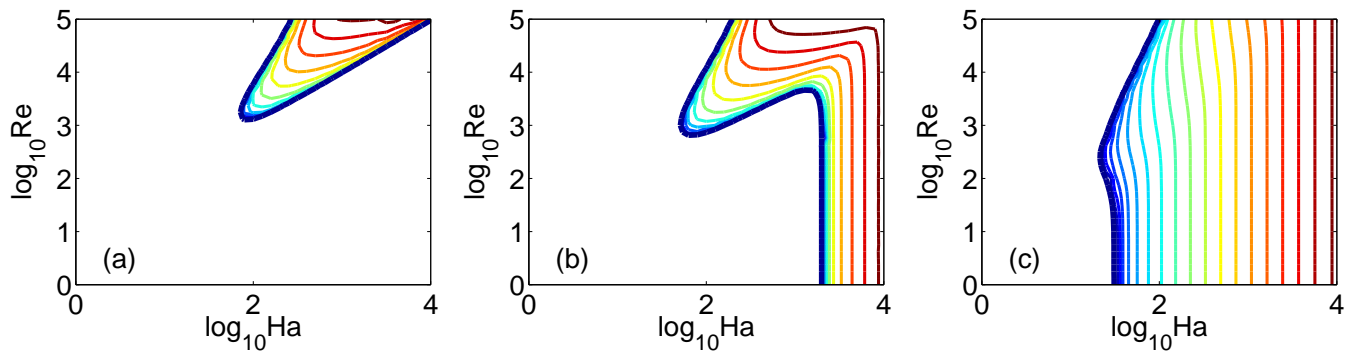}
\caption{(Color online) Instability curves for purely azimuthal magnetic fields at $\mathrm{Pm}=10^{-6}$, where $\mu_{\Omega} = 0.26, \delta=0, m=1$, with contours taken as log$_{10}\gamma$, showing the effect of changing positive $\tau$: a) $\tau=0$, b) $\tau=0.4$, and c) $\tau=100$. We can see that increasing $\tau$ causes the AMRI to be more unstable and means that the Tayler instability occurs at lower values of Ha.}
\label{fig:azitaupos}
\end{figure*}

We find that, for $\tau > 0$ at least, increasing $\tau$ makes the AMRI more unstable to both Re and Ha (Fig.\ \ref{fig:azitaupos}). The Ha threshold of instability of the Tayler instability is also affected, with more positive values of $\tau$ allowing for the Tayler instability to be excited at lower values of Ha, before reaching a limit as $\tau \rightarrow \infty$. This is not surprising, as it is well known that the Tayler instability is at its most unstable when the azimuthal field is produced solely by current in the fluid \citep{tayler1973adiabatic,seilmayer2012experimental}. The onset of the Tayler instability can be seen to rapidly approach its fully current-driven limit for values of $\tau$ as low as $\tau=5$, with a value $\tau=100$ giving an excellent approximation for solely current-driven instability.

\begin{figure}[ht]
\includegraphics[width=8.5cm]{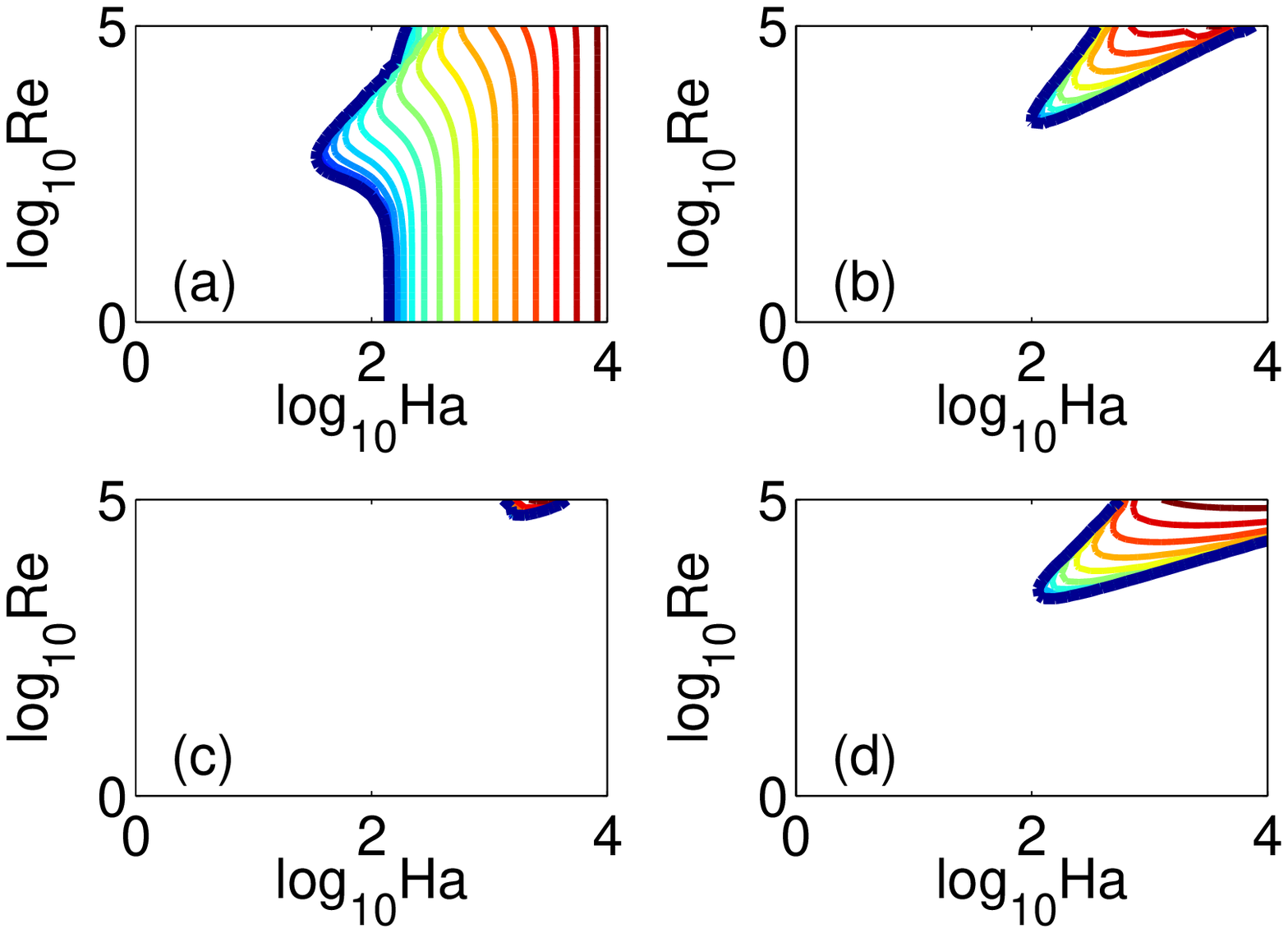}
\caption{(Color online) Instability curves for purely azimuthal magnetic fields at $\mathrm{Pm}=10^{-6}$, with contours taken for log$_{10}\gamma$, showing the effect of changing negative $\tau$, where $\mu_{\Omega}=0.26, \delta=0, m=1$. a) $\tau=-1.4$, b) $\tau=-1$, c) $\tau=-0.6$ and d) $\tau=-0.2$ Here we can see stabilisation of the AMRI around $\tau=-0.6$, whereas further decreasing $\tau$ again makes the AMRI more unstable in Re and Ha. As $|\tau|$ becomes larger the Tayler instability becomes more unstable in Ha, but at values different from that of positive $\tau$.}
\label{fig:azitauneg}
\end{figure}

Further interesting behaviour is seen when we allow $\tau$ to be negative, i.e. running current through the core and fluid in different directions (Fig.\ \ref{fig:azitauneg}). Keeping in mind the radial magnetic field profile of $B_{\phi}$, we may expect only subtle changes in the AMRI, much like for positive $\tau$ values. However, it is clear from taking negative values of $\tau$ that this is not the case at all: Between the values of $\tau = -1$ and 0 the AMRI is in fact stabilised, an effect that can be seen more greatly at $\mu_{\Omega}$ further above the Rayleigh limit. 

\begin{figure}[ht]
\centering
\includegraphics[width=8.5cm]{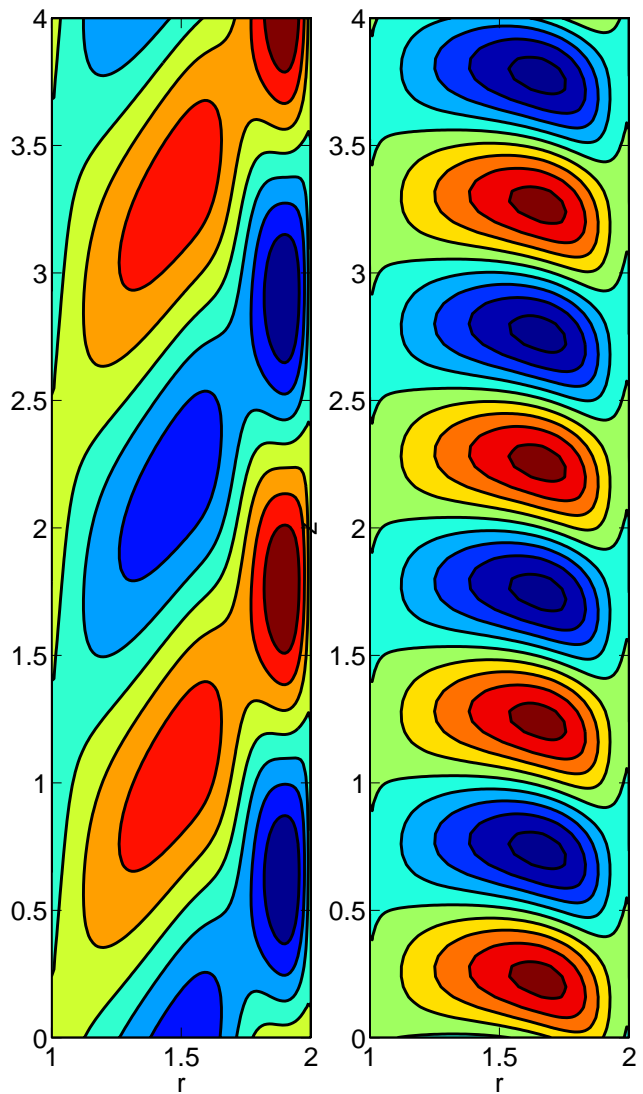}
\caption{(Color online) Eigenmodes for $\mathrm{Re}=10^4$ and $\mathrm{Ha}=10^{2.5}$ in Fig.\ \ref{fig:azitauneg} b) and d), showing the azimuthal velocity in a meridional cross section of the cylinder. Here  $\delta=0, m=1$ and $ \mu_{\Omega}=0.26$. Note that the structure of the eigenmodes for the AMRI is markedly different for $\tau$ values above and below $\tau=-0.6$, where the stabilisation of the AMRI occurred.}
\label{fig:AMRIemode}
\end{figure}

By looking at the $m=1$ eigenmodes for $\tau=-1$ and $\tau=-0.2$, (Fig.\ \ref{fig:AMRIemode}), we can see that the two modes are quite different in physical structure. In fact, it can be shown that there exists two separate branches of AMRI modes that are more unstable for different values of $\tau$. The apparent stabilisation of the AMRI around $\tau=-0.6$ can be seen to be an exchange between these modes as to which is most unstable. This phenomenon can be seen more clearly at lower values of Re when considering helical magnetic fields.

It should be noted that for more negative values of $\tau$ we see the reappearance of the Tayler instability, and as the field becomes predominantly current-driven again we see the line of marginal stability approaching its fully current-driven limit. For moderately negative $\tau$ the Tayler instability is much more difficult to excite in the Hartmann parameter than the corresponding positive $\tau$ value, likely due to unfavourable alignment of field and flow.

\subsection{Changing $\mu_{\Omega}$}

As seen in \citep{rudiger2010dissipative}, the rotation profile $\mu_{\Omega}$ can have a large effect on the stability of the AMRI and Tayler instabilities. R{\"u}diger $et$ $al$. examined cases with the steep rotation law $\mu_{\Omega}=0$, and flat rotation laws such as $\mu_{\Omega} = 0.5$. \citet{hollerbach2010nonaxisymmetric} focused attention on the Rayleigh limit up to $\mu_{\Omega}=0.35$, showing the way in which $\mu_{\Omega}$ increases the critical Reynolds number to unachievable experimental levels, but only for the case where $\tau=0$. We expand upon this to include $\mu_{\Omega}$ upwards of the Rayleigh limit for the non-zero $\tau$. We recall that it is well known that increasing $\mu_{\Omega}$ makes the AMRI more stable, however as noted by \citet{kirillov2014instabilities}, carefully changing the field profile $\tau$ can allow for instability even at large values of $\mu_{\Omega}$, though one might expect the Reynolds number required for its onset to be experimentally unachievable.

\begin{figure*}[ht]
 \centering
\includegraphics[width=\textwidth]{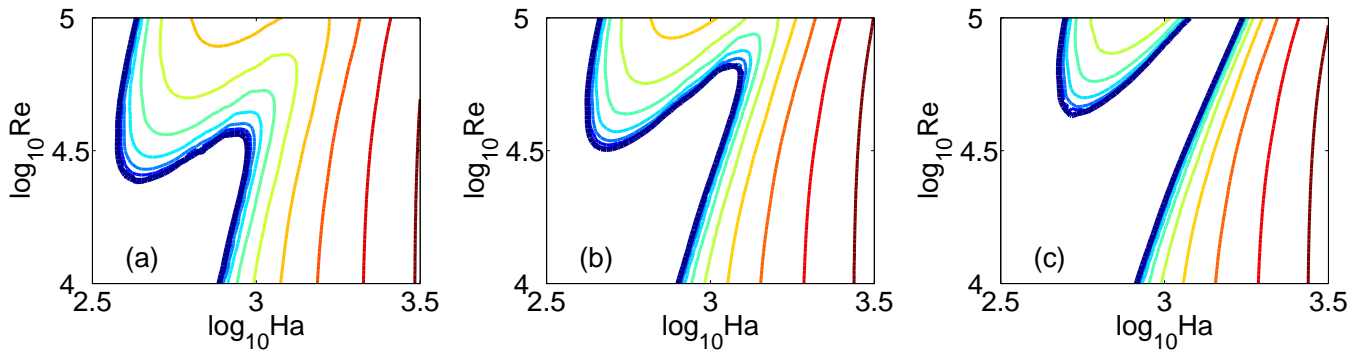}
\caption{(Color online) The $\mathrm{Pm}=10^{-6}$ analog of the weak instability seen at $\mathrm{Pm}=1$, and its dependence on $\mu_{\Omega}$, where $\tau=0.5$, $m=1$, $\delta=0$, and a) $\mu_{\Omega}=0.3975$, b) $\mu_{\Omega}=0.4$, c) $\mu_{\Omega}=0.4025$. The AMRI can clearly be seen to be present in this parameter regime, however as $\mu_{\Omega}$ is increased, greater Reynolds numbers are needed for its onset. Increasing $\mu_{\Omega}$ further such that the parameters approach the connecting curve that Kirillov identified would cause the AMRI to become completely stable.}
\label{fig:muweak}
\end{figure*}

Much like the weak instability of Chandrasekhar's equipartition solution found when benchmarking at $\mathrm{Pm}=1$, it is likely that a similarly weak instability occurs at the inductionless limit. 
Here in conjunction with $\tau$, $\mu_{\Omega}$ determines the existence of the stable equipartition region between the two instabilities, with lower values of $\mu_{\Omega}$ giving a smooth transition between the AMRI and Tayler instability. In Fig.\ \ref{fig:muweak}, which is near a point in parameter space on the Chandrasekhar line that intersects with the Liu limit connecting curve given in \cite{kirillov2014instabilities}, we vary $\mu_{\Omega}$ while keeping $\tau$ constant. We see evidence of the way in which Chandrasekhar's weak equipartition instability and corresponding stable separating region change by altering the value $\beta$, with the onset of instability becoming less attainable as $\beta$ decreases. Figure \ref{fig:muweak} also alludes to the validity of Kirillov {\it{et al.}}'s connecting curve in the linear regime through the behaviour of the AMRI, where the slight increases in $\mu_{\Omega}$  does indeed bring the regime closer to the stable region Kirillov {\it{et al.\ }}identified. It can be seen that the AMRI does become stable to linear instability as this region is entered. Note that the values $\mathrm{Ro}$ and $\mathrm{Rb}$ used by Kirillov $et$ $al$. may be expressed in terms of $\mu_{\Omega}$ and $\tau$ using $\mu_{\Omega} = (r_i/r_o)^{-2Ro}$ and $\tau = (r_i/r_o)^{-2Rb-2} - 1$. Thus, the Keplerian value of $\mathrm{Ro} = -0.75$ corresponds to $\mu_\Omega \approx 0.354$. 

Through further computations at various values of $\mu_{\Omega}$ and $\tau$ we may confirm, for at least a small section, the validity of Kirillov {\it{et al.}}'s \cite{kirillov2014instabilities} connecting curve between the upper and lower Liu limits. It can be seen that, much like in fig.\ \ref{fig:muweak}, for different values of $\mu_{\Omega}>0.25$ that adjusting $\tau$ heavily affects the stability of the AMRI, with an earlier  onset of stability, (in Re and Ha), as $\tau$ increases away from the stable region. For $\tau$ close to the connecting curve, as well as higher values of $\mu_{\Omega}$, the Re instability threshold increases rapidly, meaning that even though by increasing $\tau$ it is theoretically possible to attain the AMRI for very flat rotation profiles, the Reynolds number required eliminates the possibility of reproducing this experimentally. Finally we note that our linear stability analysis is only able to confirm a small segment of Kirillov {\it{et al.}}'s curve, between $\mathrm{Ro}=-1$ and $\mathrm{Ro}\approx{1}$. We do however show strong agreement with their WKB analysis for this region.

\subsection*{Helical magnetic fields -- $\delta\ne{0}$}

Having fully examined the sensible parameter ranges for purely azimuthal fields, we now add an axial field to the system  with relative strength given by $\delta$. Much of the previous discussion involving the AMRI in the presence of purely azimuthal magnetic fields still holds: $\tau$ and $\mu_{\Omega}$ understandably have the same effects on stability, and the results regarding Kirillov {\it{et al.}}'s connecting curve remain largely the same. The main additions here stem from separation of $\pm{m}$ due to the innate handedness of the helical field and the  stabilising or destabilising effect of the relative strength of the poloidal field. This axial field also allows for a resonance-like appearance of higher $m$ modes, as suggested by \citet{kirillov2014instabilities}. Note that though we may take $\delta$ to be negative, it is unnecessary as this is equivalent to positive $\delta$ with an oppositely signed azimuthal wave number $m$.

\subsection{Changing $\delta$}

The relative strength of the azimuthal field in the system has a large effect on the instabilities that exist. With a very weak axial component the magnetic field is still azimuthally dominated, producing the Tayler instability and nonaxisymmetric helical MRI with a particularly similar curve of marginal stability to that of a purely azimuthal field. Progressively stronger axial fields however work to either stabilise or destabilise (depending on the handedness of the flow) the Tayler instability and MRI to nonaxisymmetric instabilities. Axial fields that far outweigh the azimuthal fields simply return the SMRI, which is far more stable to differential rotation \citep{hollerbach2005}.

\begin{figure*}[ht]
 \centering
\includegraphics[width=\textwidth]{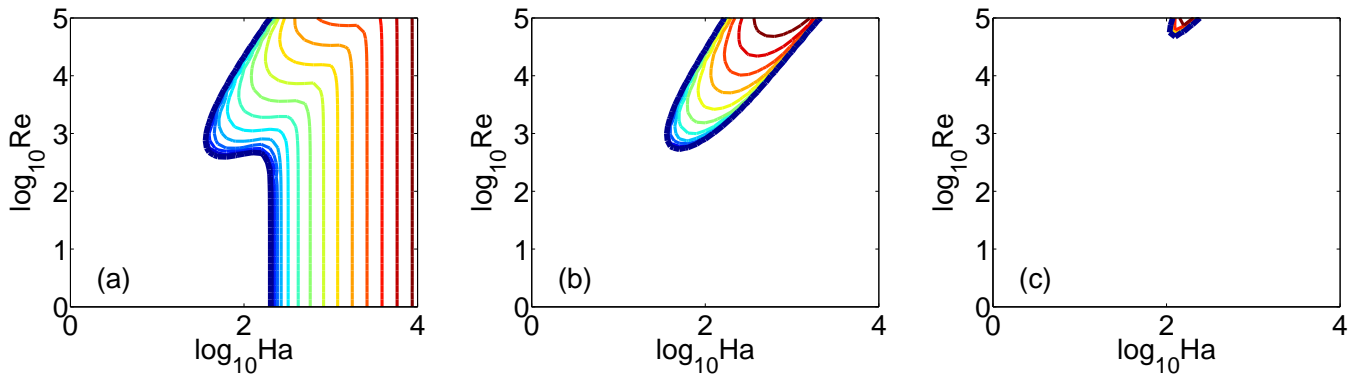}
\caption{(Color online) Instability curves for helical magnetic fields at $\mathrm{Pm}=10^{-6}$, with contours taken for log$_{10}\gamma$, showing the stabilising effect of the poloidal field at $\mu_{\Omega} = 0.26, \tau = 0.8$, $m=1$. a) $\delta=0.01$ b) $\delta=0.05$ and c) $\delta=0.5$. With a small $\delta$ there is little difference between the helical and toroidal magnetic fields, however as the field becomes more axially dominated there is an obvious stabilisation of the Tayler instability as well as the nonaxisymmetric helical MRI.}
\label{fig:deltastabilising}
\end{figure*}

For $m=1$ modes, increasing $\delta$ from zero shows this transition from weak to strongly poloidal field clearly, with obvious stabilisation of the Tayler instability at even the relatively weak $\delta = 0.05$ (Fig. \ref{fig:deltastabilising}). Even when accounting for the inherent destabilisation gained by generating the azimuthal field through current in the fluid, increasing $\delta$ by small values has a marked effect, making the nonaxisymmetric helical MRI more stable to both increasing rotation speed and total field strength. By $\delta = 0.5$, the instability is almost completely stabilised up to $\mathrm{Re}=10^{5}$, which is much greater than that attainable experimentally.

\begin{figure}[ht]
\includegraphics[width=8.5cm]{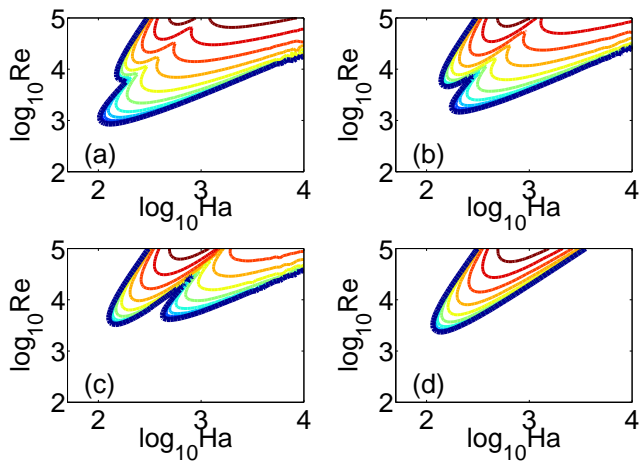}
\caption{(Color online) Instability curves at $\mathrm{Pm}=10^{-6}$ for helical magnetic fields with various azimuthal field profiles at $\mu_{\Omega} = 0.26, \delta = 0.3,\  m = -1$, with contours taken for $\log_{10}\gamma$. a) $\tau=-0.75$, b) $\tau=-0.5$, c) $\tau=-0.25$ and d) $\tau=0$. Here we can see an example of the switching between two unstable modes. It can be seen for the $m=-1$ modes that there exists two MRI modes that switch dominance for different values of $\tau$. These can be seen to have a very different eigenfunction structure (Fig. \ref{fig:2modeeig})}
\label{fig:modeswitch1}
\end{figure}

If we now consider the more interesting $m=-1$ modes, an analog to the stabilisation of the AMRI in purely azimuthal fields can be observed by examining the values $\tau$ between $-1$ and $0$  (Fig.\ \ref{fig:modeswitch1}). It appears as though the instability switches between two of its most unstable modes, which is seen more clearly at greater $|\delta|$. Due to the continuous nature of the transition between the AMRI and nonaxisymmetric helical MRI, we may assume that this is the same effect as seen in Fig.\ \ref{fig:azitauneg}. 

\begin{figure}[ht]
\centering
\includegraphics[width=8.5cm]{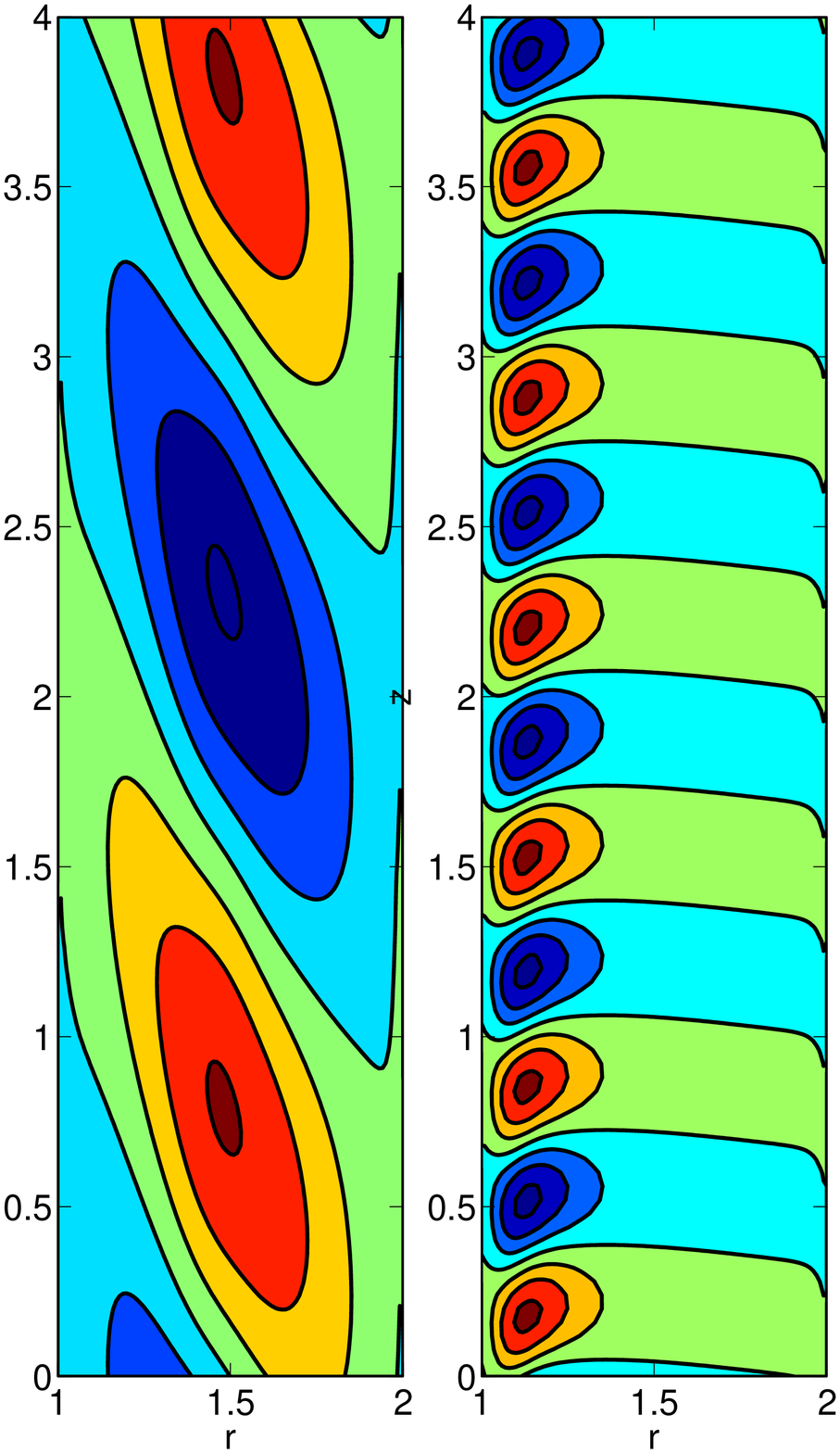}
\caption{(Color online) Plot of the two different eigenmodes from fig. \ref{fig:modeswitch1}, showing the azimuthal velocity in a meridional cross-section with $\mathrm{Re}=10^4$ and (left) $\mathrm{Ha}=10^{3}$, (right) $\mathrm{Ha}=10^{2.3}$.}
\label{fig:2modeeig}
\end{figure}

We can see that these modes are physically very different (see fig.\ \ref{fig:2modeeig}), with the left-hand mode being localised to the inner boundary and of much smaller axial wavelength, suggesting some kind of boundary mode, whereas the right-hand mode spreads over the full gap width, indicating a global mode.

\begin{figure}[ht]
 \centering
\includegraphics[width=8.5cm]{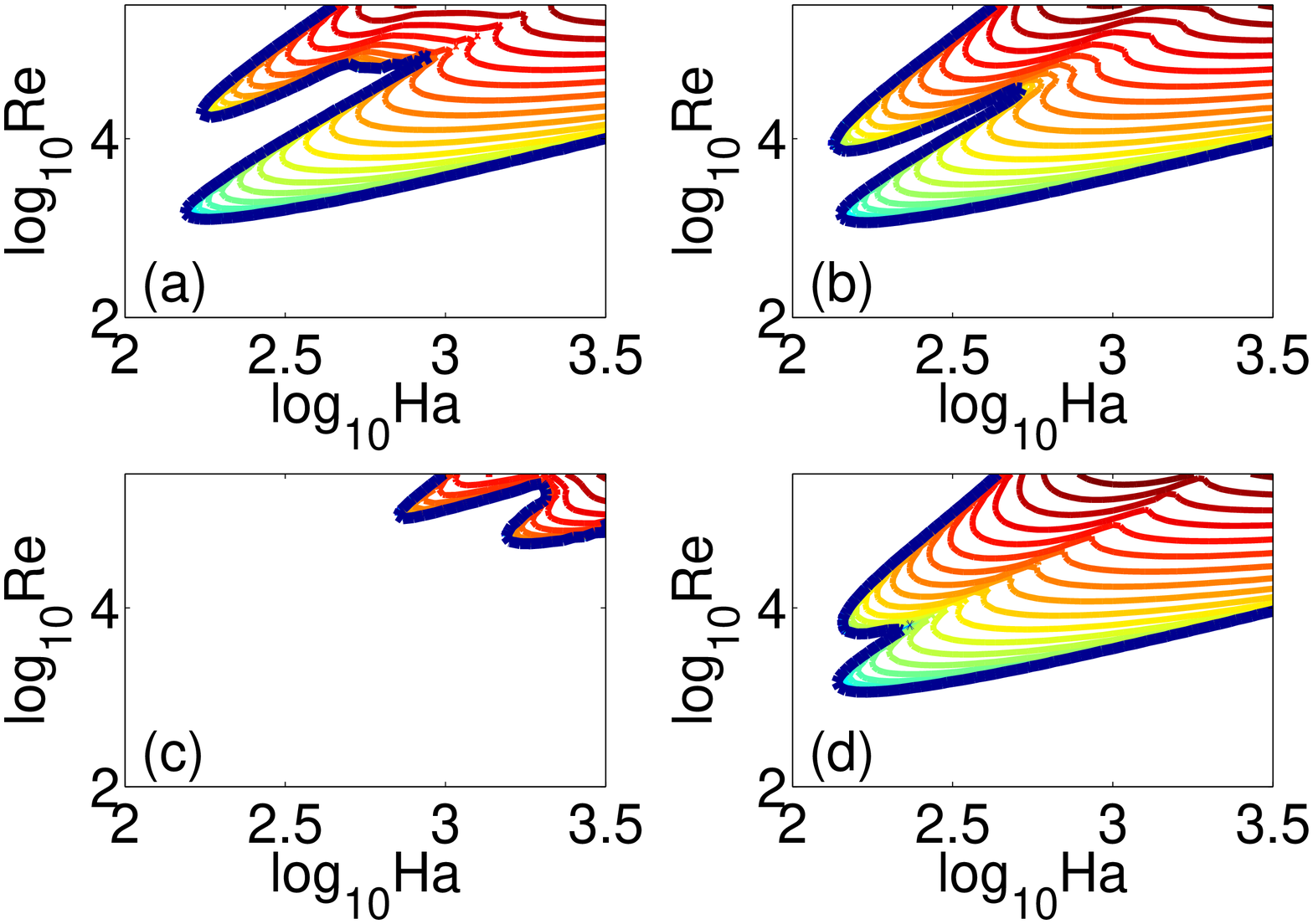}
\caption{(Color online) Instability plots at $\mathrm{Pm}=10^{-6}$ showing an inherent destabilisation caused by axial fields, with contours taken for $\log_{10}\gamma$. Here $\tau=-0.6,\mu_{\Omega}=0.26$ and $m=-1$. a) $\delta=0$, b) $\delta=0.3$, c) $\delta=0.4$ and d) $\delta=0.5$. Note that for the $m=-1$ modes the axial field can in fact destabilise the system, and that the magnitude of $\delta$ controls the separation between the two different MRI modes present. Plot (a) corresponds to the stabilisation in Fig.\ \ref{fig:azitauneg} (c), showing the existence of the two mode types mentioned.}
\label{fig:-vedelta}
\end{figure}

\begin{figure*}[ht]
 \centering
\includegraphics[width=\textwidth]{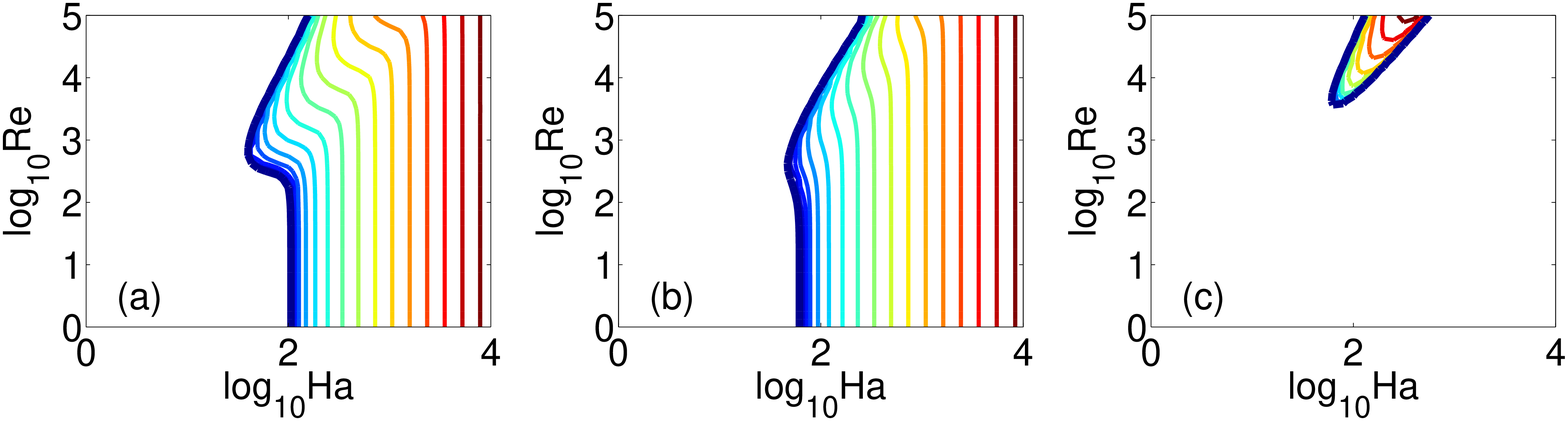}
\caption{(Color online) Contour plots at $\mathrm{Pm}=10^{-6}$ showing the symmetry breaking for $m=\pm1$ modes as $\delta$ is increased ($\mu_{\Omega}=0.26,\tau=1$). a) $\delta=0$, b) $\delta=0.3, m=-1$ and c) $\delta=0.3, m=1$. One can see that whereas the $m=1$ mode stabilises, the $m=-1$ mode becomes more unstable when a weak axial field is added.}
\label{fig:symmetry}
\end{figure*}

Now, let us consider the $m=-1$ equivalent of Fig.\ \ref{fig:deltastabilising}. By increasing $\delta$ in the range [0,0.5] we see that, whereas the $m=1$ modes would have been stabilised, the $m=-1$ modes are in fact more unstable, most likely due to favourable alignment of the flow and field directions, (Fig. \ref{fig:-vedelta}). The onset of instability can be seen at the lowest values of Ha and Re at $\delta \approx 0.3$, after which increasing the magnitude of $\delta$ again acts to stabilise the system, as one would expect. We can even see the Tayler instability, usually only excited by predominantly toroidal magnetic fields, to be present at such values of $\delta$. The profound symmetry-breaking between the $m=\pm1$ modes as $\delta$ is increased can be seen in Fig.\ \ref{fig:symmetry}.

This destabilisation by the axial field can be seen more clearly when considering specific Re (or Ha) and plotting the marginal stability in Ha (or Re) vs $\delta$. Here, motivated by the assertion by Kirillov {\it{et al.\ }}that higher modes appear predominantly in regimes with a much stronger axial than azimuthal field, we examine the stability of $m=-1,-2$ and $-3$ modes in both Re and Ha as the axial field strength is increased. We look for evidence of a banded structure, as given by Kirillov {\it{et al.}}, though due to use of a radial wave-number in their geometric optics method we cannot compare parameter regimes exactly. We focus solely on the negative $m$ modes as it can be seen, from both calculations and Kirillov {\it{et al.}}'s banded structure diagram, that positive modes are never more unstable than $m=1$, for which the onset of stability increases rapidly for relatively small values of $\delta$. In order to examine the Tayler instability and the MRI separately, the values of Re shown are taken to be $10^0$ and $10^4$, for $-100<\tau<100$, whereas when taking a fixed Ha, in order to avoid the Tayler instability a value of $\mathrm{Ha}=10^{2.4}$ was taken for $-1<\tau<1$. 

\begin{figure*}[ht]
\includegraphics[width=\textwidth]{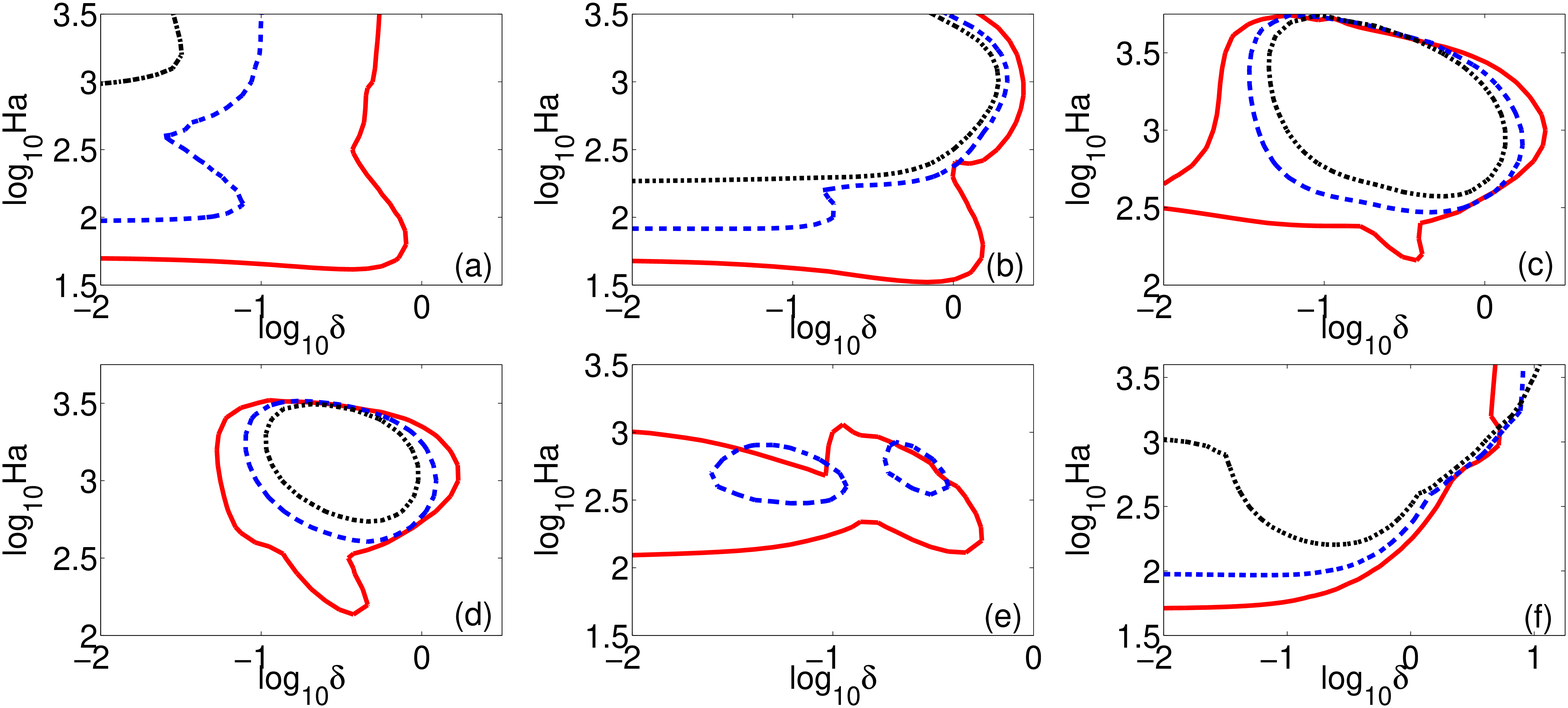}
\caption{(Color online) Marginal stability at $\mathrm{Pm}=10^{-6}$ and $\mathrm{Re}=10^4$ for various $\tau$. a) $\tau=-100$, b) $\tau=-3$, c) $\tau=-0.75$, d) $\tau-0.5$, e) $\tau=0$ and f) $\tau=100$. Plain curves correspond to $m=-1$, dashed to $m=-2$ and dot-dashed to $m=-3$. Here we can see the effect on the higher m modes that $\tau$ has. We notice that for $-3<\tau<0$ there exists a more unstable $m=-2$ mode at moderate values of $\delta$, as well as a more unstable $m=-3$ upper branch. For negative values of $\tau$, getting closer to $\tau=0$ allows for the higher $m$ modes are lesser values of $\delta$, and then for positive $\tau$ the higher $m$ modes are stable for all but relatively high values of $\delta$ for large $\tau$.}
\label{fig:higherm}
\end{figure*}

First considering $\mathrm{Re}=10^4$, we note the existence of more unstable modes for $m=-2$ and $m=-3$ (Fig. \ref{fig:higherm}), and that these are highly dependent on the value of $\tau$, much like the banded structure changed depending on Rb. For negative values of $\tau$ it is clear that the $m=-2$ and $m=-3$ modes are more unstable (to Ha) than the $m=-1$ modes at moderate values of $\delta$, on the upper and lower branches between $\delta=0.1$ and $\delta = 1$. For $\tau=0$, we may note that the upper branch of the $m=-2$ mode is the most unstable (to Ha) at similar values of $\delta$. It is at larger values of $\tau$, (positive due to the field alignment), that we see the higher $m$ modes occurring at much larger values of $\delta$, as predicted. The reliance of the existence of higher $m$ modes on $\tau$ can be justified by referring to Kirillov {\it{et al.}}'s proposed banded structure for the $m$ modes, with the value of Rb (or $\tau$) changing the 'width' in $\tau$ of each band. Indeed, for the negative values of $\tau$ given here the banded nature of the mode number $m$ can be seen in the $m=-1,0,-2,-1,...$ pattern given by Kirillov {\it{et al{.\ }}}(if one removes the $m=0$ HMRI mode). The same overall dependences can be seen for the Tayler instability when taking $\mathrm{Re}=1$. 

When considering the Reynolds stability as $\delta$ is varied, we note that, aside from a small range around $\tau=-1$, it is the $m=-1$ mode that is most unstable for all values $\delta$. This is interesting in that it suggests that, while more unstable $m=-2,-3$ modes do appear at higher $\delta$ in a resonant fashion, they are more unstable to increases in the magnetic field strength, not rotation rate. This suggests that increasing the magnetic field strength drives the higher modes comparatively more than $m=-1$, whereas the rotation favours no mode in particular.

\section{Conclusion}
In the present work we have fully explored the relevant parameter space for the nonaxisymmetric MRI variants, highlighting a number of instability results as well as confirming others in the linear regime. We have shown clearly the large effect that the steepness of the  magnetic field profile has on the onset of stability for both the MRI and Tayler instability, with fields generated by running current in the fluid allowing for earlier onset of marginal instability, and the effect this has on the flow structure of the non-axisymmetric helical MRI modes, showing in fig.\ \ref{fig:2modeeig} the two different types of modes that may be excited. We have added to previous work concerning Chandrasekhar's equipartition, showing clearly the comparative weakness in the growth of this instability as well as giving an alternative example. Further to this, we have shown evidence of the location of sections of Kirillov {\it{et al.}}'s connecting curve between the Liu limits in the linear regime, and have shown the existence of the higher $m$ modes previously predicted for axially dominated fields, offering confirmation to the validity of their WKB analysis in the linear regime, as well as the `banded' nature of the modes.

We identify a number of possible avenues for further work. In particular, for the two different modes structures present in Fig.\ \ref{fig:2modeeig}, the question remains as to why each of the two modes is dominant for either side of $\tau\approx{-0.6}$, and what significance, if any, the value $\tau\approx-0.6$, (at which both modes are stabilised in a purely toroidal field), holds. A full asymptotic analysis of our perturbation equations around this parameter range may prove to be fruitful in exploring this question. The regime $\mu_\Omega > 1$ also seems to yield some interesting results \cite{rudiger2015subcritical,stefani2015destabilization}, and may warrant further investigation.

\section{Acknowledgements} 
AC was supported by an STFC studentship; RH was supported by STFC Grant No. ST/K000853/1. 

We thank Dr Frank Stefani, as well as the anonymous referees, for their helpful suggestions and comments.


\begin{thebibliography}{31}%
\makeatletter
\providecommand \@ifxundefined [1]{%
 \@ifx{#1\undefined}
}%
\providecommand \@ifnum [1]{%
 \ifnum #1\expandafter \@firstoftwo
 \else \expandafter \@secondoftwo
 \fi
}%
\providecommand \@ifx [1]{%
 \ifx #1\expandafter \@firstoftwo
 \else \expandafter \@secondoftwo
 \fi
}%
\providecommand \natexlab [1]{#1}%
\providecommand \enquote  [1]{``#1''}%
\providecommand \bibnamefont  [1]{#1}%
\providecommand \bibfnamefont [1]{#1}%
\providecommand \citenamefont [1]{#1}%
\providecommand \href@noop [0]{\@secondoftwo}%
\providecommand \href [0]{\begingroup \@sanitize@url \@href}%
\providecommand \@href[1]{\@@startlink{#1}\@@href}%
\providecommand \@@href[1]{\endgroup#1\@@endlink}%
\providecommand \@sanitize@url [0]{\catcode `\\12\catcode `\$12\catcode
  `\&12\catcode `\#12\catcode `\^12\catcode `\_12\catcode `\%12\relax}%
\providecommand \@@startlink[1]{}%
\providecommand \@@endlink[0]{}%
\providecommand \url  [0]{\begingroup\@sanitize@url \@url }%
\providecommand \@url [1]{\endgroup\@href {#1}{\urlprefix }}%
\providecommand \urlprefix  [0]{URL }%
\providecommand \Eprint [0]{\href }%
\providecommand \doibase [0]{http://dx.doi.org/}%
\providecommand \selectlanguage [0]{\@gobble}%
\providecommand \bibinfo  [0]{\@secondoftwo}%
\providecommand \bibfield  [0]{\@secondoftwo}%
\providecommand \translation [1]{[#1]}%
\providecommand \BibitemOpen [0]{}%
\providecommand \bibitemStop [0]{}%
\providecommand \bibitemNoStop [0]{.\EOS\space}%
\providecommand \EOS [0]{\spacefactor3000\relax}%
\providecommand \BibitemShut  [1]{\csname bibitem#1\endcsname}%
\let\auto@bib@innerbib\@empty
\bibitem [{\citenamefont {Velikhov}(1959)}]{velikhov_1959}%
  \BibitemOpen
  \bibfield  {author} {\bibinfo {author} {\bibfnamefont {E.}~\bibnamefont
  {Velikhov}},\ }\href@noop {} {\bibfield  {journal} {\bibinfo  {journal} {Sov.
  Phys. JETP}\ }\textbf {\bibinfo {volume} {36}},\ \bibinfo {pages} {995}
  (\bibinfo {year} {1959})}\BibitemShut {NoStop}%
\bibitem [{\citenamefont {Balbus}\ and\ \citenamefont
  {Hawley}(1991)}]{Balbus1991}%
  \BibitemOpen
  \bibfield  {author} {\bibinfo {author} {\bibfnamefont {S.~A.}\ \bibnamefont
  {Balbus}}\ and\ \bibinfo {author} {\bibfnamefont {J.~F.}\ \bibnamefont
  {Hawley}},\ }\href@noop {} {\bibfield  {journal} {\bibinfo  {journal}
  {Astrophys. J.}\ }\textbf {\bibinfo {volume} {376}},\ \bibinfo {pages} {214}
  (\bibinfo {year} {1991})}\BibitemShut {NoStop}%
\bibitem [{\citenamefont {Julien}\ and\ \citenamefont
  {Knobloch}(2010)}]{julien2010magnetorotational}%
  \BibitemOpen
  \bibfield  {author} {\bibinfo {author} {\bibfnamefont {K.}~\bibnamefont
  {Julien}}\ and\ \bibinfo {author} {\bibfnamefont {E.}~\bibnamefont
  {Knobloch}},\ }\href@noop {} {\bibfield  {journal} {\bibinfo  {journal}
  {Phil. Trans. Royal. Soc. A}\ }\textbf {\bibinfo {volume} {368}},\ \bibinfo
  {pages} {1607} (\bibinfo {year} {2010})}\BibitemShut {NoStop}%
\bibitem [{\citenamefont {Vandakurov}(1972)}]{vandakurov1972theory}%
  \BibitemOpen
  \bibfield  {author} {\bibinfo {author} {\bibfnamefont {Y.~V.}\ \bibnamefont
  {Vandakurov}},\ }\href@noop {} {\bibfield  {journal} {\bibinfo  {journal}
  {Sov. Astron.}\ }\textbf {\bibinfo {volume} {16}},\ \bibinfo {pages} {265}
  (\bibinfo {year} {1972})}\BibitemShut {NoStop}%
\bibitem [{\citenamefont {Tayler}(1973)}]{tayler1973adiabatic}%
  \BibitemOpen
  \bibfield  {author} {\bibinfo {author} {\bibfnamefont {R.}~\bibnamefont
  {Tayler}},\ }\href@noop {} {\bibfield  {journal} {\bibinfo  {journal} {Mon.
  Not. R. Astron. Soc.}\ }\textbf {\bibinfo {volume} {161}},\ \bibinfo {pages}
  {365} (\bibinfo {year} {1973})}\BibitemShut {NoStop}%
\bibitem [{\citenamefont {Spruit}(2002)}]{spruit2002dynamo}%
  \BibitemOpen
  \bibfield  {author} {\bibinfo {author} {\bibfnamefont {H.}~\bibnamefont
  {Spruit}},\ }\href@noop {} {\bibfield  {journal} {\bibinfo  {journal}
  {Astron. Astrophys.}\ }\textbf {\bibinfo {volume} {381}},\ \bibinfo {pages}
  {923} (\bibinfo {year} {2002})}\BibitemShut {NoStop}%
\bibitem [{\citenamefont {Weber}\ \emph {et~al.}(2015)\citenamefont {Weber},
  \citenamefont {Galindo}, \citenamefont {Priede}, \citenamefont {Stefani},\
  and\ \citenamefont {Weier}}]{weber2015influence}%
  \BibitemOpen
  \bibfield  {author} {\bibinfo {author} {\bibfnamefont {N.}~\bibnamefont
  {Weber}}, \bibinfo {author} {\bibfnamefont {V.}~\bibnamefont {Galindo}},
  \bibinfo {author} {\bibfnamefont {J.}~\bibnamefont {Priede}}, \bibinfo
  {author} {\bibfnamefont {F.}~\bibnamefont {Stefani}}, \ and\ \bibinfo
  {author} {\bibfnamefont {T.}~\bibnamefont {Weier}},\ }\href@noop {}
  {\bibfield  {journal} {\bibinfo  {journal} {Phys. Fluids}\ }\textbf {\bibinfo
  {volume} {27}},\ \bibinfo {pages} {014103} (\bibinfo {year}
  {2015})}\BibitemShut {NoStop}%
\bibitem [{\citenamefont {Herreman}\ \emph {et~al.}(2015)\citenamefont
  {Herreman}, \citenamefont {Nore}, \citenamefont {Cappanera},\ and\
  \citenamefont {Guermond}}]{herreman2015tayler}%
  \BibitemOpen
  \bibfield  {author} {\bibinfo {author} {\bibfnamefont {W.}~\bibnamefont
  {Herreman}}, \bibinfo {author} {\bibfnamefont {C.}~\bibnamefont {Nore}},
  \bibinfo {author} {\bibfnamefont {L.}~\bibnamefont {Cappanera}}, \ and\
  \bibinfo {author} {\bibfnamefont {J.-L.}\ \bibnamefont {Guermond}},\
  }\href@noop {} {\bibfield  {journal} {\bibinfo  {journal} {J. Fluid. Mech.}\
  }\textbf {\bibinfo {volume} {771}},\ \bibinfo {pages} {79} (\bibinfo {year}
  {2015})}\BibitemShut {NoStop}%
\bibitem [{\citenamefont {R{\"u}diger}\ and\ \citenamefont
  {Zhang}(2001)}]{rudiger2001mhd}%
  \BibitemOpen
  \bibfield  {author} {\bibinfo {author} {\bibfnamefont {G.}~\bibnamefont
  {R{\"u}diger}}\ and\ \bibinfo {author} {\bibfnamefont {Y.}~\bibnamefont
  {Zhang}},\ }\href@noop {} {\bibfield  {journal} {\bibinfo  {journal} {Astro.
  \& Astrophys.}\ }\textbf {\bibinfo {volume} {378}},\ \bibinfo {pages} {302}
  (\bibinfo {year} {2001})}\BibitemShut {NoStop}%
\bibitem [{\citenamefont {Goodman}\ and\ \citenamefont
  {Ji}(2002)}]{goodman2002}%
  \BibitemOpen
  \bibfield  {author} {\bibinfo {author} {\bibfnamefont {J.}~\bibnamefont
  {Goodman}}\ and\ \bibinfo {author} {\bibfnamefont {H.}~\bibnamefont {Ji}},\
  }\href@noop {} {\bibfield  {journal} {\bibinfo  {journal} {J. Fluid Mech.}\
  }\textbf {\bibinfo {volume} {462}},\ \bibinfo {pages} {365} (\bibinfo {year}
  {2002})}\BibitemShut {NoStop}%
\bibitem [{\citenamefont {{Hollerbach}}\ and\ \citenamefont
  {{Fournier}}(2004)}]{2004hollerbach}%
  \BibitemOpen
  \bibfield  {author} {\bibinfo {author} {\bibfnamefont {R.}~\bibnamefont
  {{Hollerbach}}}\ and\ \bibinfo {author} {\bibfnamefont {A.}~\bibnamefont
  {{Fournier}}},\ }\href {\doibase 10.1063/1.1832141} {\emph {\bibinfo {title}
  {MHD Couette Flows: Experiments and Models}}},\ edited by\ \bibinfo {editor}
  {\bibfnamefont {R.}~\bibnamefont {{Rosner}}}, \bibinfo {editor}
  {\bibfnamefont {G.}~\bibnamefont {{R{\"u}diger}}}, \ and\ \bibinfo {editor}
  {\bibfnamefont {A.}~\bibnamefont {{Bonanno}}},\ \bibinfo {series} {American
  Institute of Physics Conference Series}, Vol.\ \bibinfo {volume} {733}\
  (\bibinfo {year} {2004})\ pp.\ \bibinfo {pages} {114--121},\ \Eprint
  {http://arxiv.org/abs/astro-ph/0506081} {astro-ph/0506081} \BibitemShut
  {NoStop}%
\bibitem [{\citenamefont {Nornberg}\ \emph {et~al.}(2010)\citenamefont
  {Nornberg}, \citenamefont {Ji}, \citenamefont {Schartman}, \citenamefont
  {Roach},\ and\ \citenamefont {Goodman}}]{nornberg2010observation}%
  \BibitemOpen
  \bibfield  {author} {\bibinfo {author} {\bibfnamefont {M.}~\bibnamefont
  {Nornberg}}, \bibinfo {author} {\bibfnamefont {H.}~\bibnamefont {Ji}},
  \bibinfo {author} {\bibfnamefont {E.}~\bibnamefont {Schartman}}, \bibinfo
  {author} {\bibfnamefont {A.}~\bibnamefont {Roach}}, \ and\ \bibinfo {author}
  {\bibfnamefont {J.}~\bibnamefont {Goodman}},\ }\href@noop {} {\bibfield
  {journal} {\bibinfo  {journal} {Phys. Rev. Lett.}\ }\textbf {\bibinfo
  {volume} {104}},\ \bibinfo {pages} {074501} (\bibinfo {year}
  {2010})}\BibitemShut {NoStop}%
\bibitem [{\citenamefont {Hollerbach}\ and\ \citenamefont
  {R{\"u}diger}(2005)}]{hollerbach2005}%
  \BibitemOpen
  \bibfield  {author} {\bibinfo {author} {\bibfnamefont {R.}~\bibnamefont
  {Hollerbach}}\ and\ \bibinfo {author} {\bibfnamefont {G.}~\bibnamefont
  {R{\"u}diger}},\ }\href@noop {} {\bibfield  {journal} {\bibinfo  {journal}
  {Phys. Rev. Lett.}\ }\textbf {\bibinfo {volume} {95}},\ \bibinfo {pages}
  {124501} (\bibinfo {year} {2005})}\BibitemShut {NoStop}%
\bibitem [{\citenamefont {Stefani}\ \emph {et~al.}(2006)\citenamefont
  {Stefani}, \citenamefont {Gundrum}, \citenamefont {Gerbeth}, \citenamefont
  {R{\"u}diger}, \citenamefont {Schultz}, \citenamefont {Szklarski},\ and\
  \citenamefont {Hollerbach}}]{stefani2006experimental}%
  \BibitemOpen
  \bibfield  {author} {\bibinfo {author} {\bibfnamefont {F.}~\bibnamefont
  {Stefani}}, \bibinfo {author} {\bibfnamefont {T.}~\bibnamefont {Gundrum}},
  \bibinfo {author} {\bibfnamefont {G.}~\bibnamefont {Gerbeth}}, \bibinfo
  {author} {\bibfnamefont {G.}~\bibnamefont {R{\"u}diger}}, \bibinfo {author}
  {\bibfnamefont {M.}~\bibnamefont {Schultz}}, \bibinfo {author} {\bibfnamefont
  {J.}~\bibnamefont {Szklarski}}, \ and\ \bibinfo {author} {\bibfnamefont
  {R.}~\bibnamefont {Hollerbach}},\ }\href@noop {} {\bibfield  {journal}
  {\bibinfo  {journal} {Phys. Rev. Lett.}\ }\textbf {\bibinfo {volume} {97}},\
  \bibinfo {pages} {184502} (\bibinfo {year} {2006})}\BibitemShut {NoStop}%
\bibitem [{\citenamefont {Stefani}\ \emph {et~al.}(2009)\citenamefont
  {Stefani}, \citenamefont {Gerbeth}, \citenamefont {Gundrum}, \citenamefont
  {Hollerbach}, \citenamefont {Priede}, \citenamefont {R{\"u}diger},\ and\
  \citenamefont {Szklarski}}]{stefani2009helical}%
  \BibitemOpen
  \bibfield  {author} {\bibinfo {author} {\bibfnamefont {F.}~\bibnamefont
  {Stefani}}, \bibinfo {author} {\bibfnamefont {G.}~\bibnamefont {Gerbeth}},
  \bibinfo {author} {\bibfnamefont {T.}~\bibnamefont {Gundrum}}, \bibinfo
  {author} {\bibfnamefont {R.}~\bibnamefont {Hollerbach}}, \bibinfo {author}
  {\bibfnamefont {J.}~\bibnamefont {Priede}}, \bibinfo {author} {\bibfnamefont
  {G.}~\bibnamefont {R{\"u}diger}}, \ and\ \bibinfo {author} {\bibfnamefont
  {J.}~\bibnamefont {Szklarski}},\ }\href@noop {} {\bibfield  {journal}
  {\bibinfo  {journal} {Phys. Rev. E}\ }\textbf {\bibinfo {volume} {80}},\
  \bibinfo {pages} {066303} (\bibinfo {year} {2009})}\BibitemShut {NoStop}%
\bibitem [{\citenamefont {Priede}\ and\ \citenamefont
  {Gerbeth}(2009)}]{priede2009absolute}%
  \BibitemOpen
  \bibfield  {author} {\bibinfo {author} {\bibfnamefont {J.}~\bibnamefont
  {Priede}}\ and\ \bibinfo {author} {\bibfnamefont {G.}~\bibnamefont
  {Gerbeth}},\ }\href@noop {} {\bibfield  {journal} {\bibinfo  {journal} {Phys.
  Rev. E}\ }\textbf {\bibinfo {volume} {79}},\ \bibinfo {pages} {046310}
  (\bibinfo {year} {2009})}\BibitemShut {NoStop}%
\bibitem [{\citenamefont {Ogilvie}\ and\ \citenamefont
  {Pringle}(1996)}]{ogilvie1996non}%
  \BibitemOpen
  \bibfield  {author} {\bibinfo {author} {\bibfnamefont {G.}~\bibnamefont
  {Ogilvie}}\ and\ \bibinfo {author} {\bibfnamefont {J.}~\bibnamefont
  {Pringle}},\ }\href@noop {} {\bibfield  {journal} {\bibinfo  {journal} {Mon.
  Not. R. Astron. Soc.}\ }\textbf {\bibinfo {volume} {279}},\ \bibinfo {pages}
  {152} (\bibinfo {year} {1996})}\BibitemShut {NoStop}%
\bibitem [{\citenamefont {Hollerbach}\ \emph {et~al.}(2010)\citenamefont
  {Hollerbach}, \citenamefont {Teeluck},\ and\ \citenamefont
  {R{\"u}diger}}]{hollerbach2010nonaxisymmetric}%
  \BibitemOpen
  \bibfield  {author} {\bibinfo {author} {\bibfnamefont {R.}~\bibnamefont
  {Hollerbach}}, \bibinfo {author} {\bibfnamefont {V.}~\bibnamefont {Teeluck}},
  \ and\ \bibinfo {author} {\bibfnamefont {G.}~\bibnamefont {R{\"u}diger}},\
  }\href@noop {} {\bibfield  {journal} {\bibinfo  {journal} {Phys. Rev. Lett.}\
  }\textbf {\bibinfo {volume} {104}},\ \bibinfo {pages} {044502} (\bibinfo
  {year} {2010})}\BibitemShut {NoStop}%
\bibitem [{\citenamefont {Seilmayer}\ \emph {et~al.}(2014)\citenamefont
  {Seilmayer}, \citenamefont {Galindo}, \citenamefont {Gerbeth}, \citenamefont
  {Gundrum}, \citenamefont {Stefani}, \citenamefont {Gellert}, \citenamefont
  {R\"udiger}, \citenamefont {Schultz},\ and\ \citenamefont
  {Hollerbach}}]{seilmayer2014experimental}%
  \BibitemOpen
  \bibfield  {author} {\bibinfo {author} {\bibfnamefont {M.}~\bibnamefont
  {Seilmayer}}, \bibinfo {author} {\bibfnamefont {V.}~\bibnamefont {Galindo}},
  \bibinfo {author} {\bibfnamefont {G.}~\bibnamefont {Gerbeth}}, \bibinfo
  {author} {\bibfnamefont {T.}~\bibnamefont {Gundrum}}, \bibinfo {author}
  {\bibfnamefont {F.}~\bibnamefont {Stefani}}, \bibinfo {author} {\bibfnamefont
  {M.}~\bibnamefont {Gellert}}, \bibinfo {author} {\bibfnamefont
  {G.}~\bibnamefont {R\"udiger}}, \bibinfo {author} {\bibfnamefont
  {M.}~\bibnamefont {Schultz}}, \ and\ \bibinfo {author} {\bibfnamefont
  {R.}~\bibnamefont {Hollerbach}},\ }\href {\doibase
  10.1103/PhysRevLett.113.024505} {\bibfield  {journal} {\bibinfo  {journal}
  {Phys. Rev. Lett.}\ }\textbf {\bibinfo {volume} {113}},\ \bibinfo {pages}
  {024505} (\bibinfo {year} {2014})}\BibitemShut {NoStop}%
\bibitem [{\citenamefont {Kirillov}\ \emph {et~al.}(2014)\citenamefont
  {Kirillov}, \citenamefont {Stefani},\ and\ \citenamefont
  {Fukumoto}}]{kirillov2014instabilities}%
  \BibitemOpen
  \bibfield  {author} {\bibinfo {author} {\bibfnamefont {O.~N.}\ \bibnamefont
  {Kirillov}}, \bibinfo {author} {\bibfnamefont {F.}~\bibnamefont {Stefani}}, \
  and\ \bibinfo {author} {\bibfnamefont {Y.}~\bibnamefont {Fukumoto}},\
  }\href@noop {} {\bibfield  {journal} {\bibinfo  {journal} {J. Fluid Mech.}\
  }\textbf {\bibinfo {volume} {760}},\ \bibinfo {pages} {591} (\bibinfo {year}
  {2014})}\BibitemShut {NoStop}%
\bibitem [{\citenamefont {Liu}\ \emph {et~al.}(2006)\citenamefont {Liu},
  \citenamefont {Goodman}, \citenamefont {Herron},\ and\ \citenamefont
  {Ji}}]{liu2006}%
  \BibitemOpen
  \bibfield  {author} {\bibinfo {author} {\bibfnamefont {W.}~\bibnamefont
  {Liu}}, \bibinfo {author} {\bibfnamefont {J.}~\bibnamefont {Goodman}},
  \bibinfo {author} {\bibfnamefont {I.}~\bibnamefont {Herron}}, \ and\ \bibinfo
  {author} {\bibfnamefont {H.}~\bibnamefont {Ji}},\ }\href {\doibase
  10.1103/PhysRevE.74.056302} {\bibfield  {journal} {\bibinfo  {journal} {Phys.
  Rev. E}\ }\textbf {\bibinfo {volume} {74}},\ \bibinfo {pages} {056302}
  (\bibinfo {year} {2006})}\BibitemShut {NoStop}%
\bibitem [{\citenamefont {Priede}(2011)}]{priede2011inviscid}%
  \BibitemOpen
  \bibfield  {author} {\bibinfo {author} {\bibfnamefont {J.}~\bibnamefont
  {Priede}},\ }\href@noop {} {\bibfield  {journal} {\bibinfo  {journal} {Phys.
  Rev. E}\ }\textbf {\bibinfo {volume} {84}},\ \bibinfo {pages} {066314}
  (\bibinfo {year} {2011})}\BibitemShut {NoStop}%
\bibitem [{\citenamefont {Kirillov}\ and\ \citenamefont
  {Stefani}(2013)}]{kirillov2013extending}%
  \BibitemOpen
  \bibfield  {author} {\bibinfo {author} {\bibfnamefont {O.~N.}\ \bibnamefont
  {Kirillov}}\ and\ \bibinfo {author} {\bibfnamefont {F.}~\bibnamefont
  {Stefani}},\ }\href@noop {} {\bibfield  {journal} {\bibinfo  {journal} {Phys.
  Rev. Lett.}\ }\textbf {\bibinfo {volume} {111}},\ \bibinfo {pages} {061103}
  (\bibinfo {year} {2013})}\BibitemShut {NoStop}%
\bibitem [{\citenamefont {R{\"u}diger}\ \emph {et~al.}(2010)\citenamefont
  {R{\"u}diger}, \citenamefont {Gellert}, \citenamefont {Schultz},\ and\
  \citenamefont {Hollerbach}}]{rudiger2010dissipative}%
  \BibitemOpen
  \bibfield  {author} {\bibinfo {author} {\bibfnamefont {G.}~\bibnamefont
  {R{\"u}diger}}, \bibinfo {author} {\bibfnamefont {M.}~\bibnamefont
  {Gellert}}, \bibinfo {author} {\bibfnamefont {M.}~\bibnamefont {Schultz}}, \
  and\ \bibinfo {author} {\bibfnamefont {R.}~\bibnamefont {Hollerbach}},\
  }\href@noop {} {\bibfield  {journal} {\bibinfo  {journal} {Phys. Rev. E}\
  }\textbf {\bibinfo {volume} {82}},\ \bibinfo {pages} {016319} (\bibinfo
  {year} {2010})}\BibitemShut {NoStop}%
\bibitem [{\citenamefont {Priede}(2015)}]{priede2015metamorphosis}%
  \BibitemOpen
  \bibfield  {author} {\bibinfo {author} {\bibfnamefont {J.}~\bibnamefont
  {Priede}},\ }\href {\doibase 10.1103/PhysRevE.91.033014} {\bibfield
  {journal} {\bibinfo  {journal} {Phys. Rev. E}\ }\textbf {\bibinfo {volume}
  {91}},\ \bibinfo {pages} {033014} (\bibinfo {year} {2015})}\BibitemShut
  {NoStop}%
\bibitem [{\citenamefont {R{\"u}diger}\ \emph {et~al.}(2014)\citenamefont
  {R{\"u}diger}, \citenamefont {Schultz}, \citenamefont {Stefani},\ and\
  \citenamefont {Mond}}]{rudiger2014diffusive}%
  \BibitemOpen
  \bibfield  {author} {\bibinfo {author} {\bibfnamefont {G.}~\bibnamefont
  {R{\"u}diger}}, \bibinfo {author} {\bibfnamefont {M.}~\bibnamefont
  {Schultz}}, \bibinfo {author} {\bibfnamefont {F.}~\bibnamefont {Stefani}}, \
  and\ \bibinfo {author} {\bibfnamefont {M.}~\bibnamefont {Mond}},\ }\href@noop
  {} {\bibfield  {journal} {\bibinfo  {journal} {arXiv preprint
  arXiv:1407.0240}\ } (\bibinfo {year} {2014})}\BibitemShut {NoStop}%
\bibitem [{\citenamefont {Chandrasekhar}(1956)}]{chandrasekhar1956stability}%
  \BibitemOpen
  \bibfield  {author} {\bibinfo {author} {\bibfnamefont {S.}~\bibnamefont
  {Chandrasekhar}},\ }\href@noop {} {\bibfield  {journal} {\bibinfo  {journal}
  {Proc. Nat. Acad. Sci.}\ }\textbf {\bibinfo {volume} {42}},\ \bibinfo {pages}
  {273} (\bibinfo {year} {1956})}\BibitemShut {NoStop}%
\bibitem [{\citenamefont {Tataronis}\ and\ \citenamefont
  {Mond}(1987)}]{mond1987}%
  \BibitemOpen
  \bibfield  {author} {\bibinfo {author} {\bibfnamefont {J.~A.}\ \bibnamefont
  {Tataronis}}\ and\ \bibinfo {author} {\bibfnamefont {M.}~\bibnamefont
  {Mond}},\ }\href {\doibase http://dx.doi.org/10.1063/1.866064} {\bibfield
  {journal} {\bibinfo  {journal} {Phys. Fluids}\ }\textbf {\bibinfo {volume}
  {30}},\ \bibinfo {pages} {84} (\bibinfo {year} {1987})}\BibitemShut {NoStop}%
\bibitem [{\citenamefont {Seilmayer}\ \emph {et~al.}(2012)\citenamefont
  {Seilmayer}, \citenamefont {Stefani}, \citenamefont {Gundrum}, \citenamefont
  {Weier}, \citenamefont {Gerbeth}, \citenamefont {Gellert},\ and\
  \citenamefont {R\"udiger}}]{seilmayer2012experimental}%
  \BibitemOpen
  \bibfield  {author} {\bibinfo {author} {\bibfnamefont {M.}~\bibnamefont
  {Seilmayer}}, \bibinfo {author} {\bibfnamefont {F.}~\bibnamefont {Stefani}},
  \bibinfo {author} {\bibfnamefont {T.}~\bibnamefont {Gundrum}}, \bibinfo
  {author} {\bibfnamefont {T.}~\bibnamefont {Weier}}, \bibinfo {author}
  {\bibfnamefont {G.}~\bibnamefont {Gerbeth}}, \bibinfo {author} {\bibfnamefont
  {M.}~\bibnamefont {Gellert}}, \ and\ \bibinfo {author} {\bibfnamefont
  {G.}~\bibnamefont {R\"udiger}},\ }\href {\doibase
  10.1103/PhysRevLett.108.244501} {\bibfield  {journal} {\bibinfo  {journal}
  {Phys. Rev. Lett.}\ }\textbf {\bibinfo {volume} {108}},\ \bibinfo {pages}
  {244501} (\bibinfo {year} {2012})}\BibitemShut {NoStop}%
\bibitem [{\citenamefont {R{\"u}diger}\ \emph {et~al.}(2015)\citenamefont
  {R{\"u}diger}, \citenamefont {Schultz}, \citenamefont {Gellert},\ and\
  \citenamefont {Stefani}}]{rudiger2015subcritical}%
  \BibitemOpen
  \bibfield  {author} {\bibinfo {author} {\bibfnamefont {G.}~\bibnamefont
  {R{\"u}diger}}, \bibinfo {author} {\bibfnamefont {M.}~\bibnamefont
  {Schultz}}, \bibinfo {author} {\bibfnamefont {M.}~\bibnamefont {Gellert}}, \
  and\ \bibinfo {author} {\bibfnamefont {F.}~\bibnamefont {Stefani}},\
  }\href@noop {} {\bibfield  {journal} {\bibinfo  {journal} {arXiv preprint
  arXiv:1505.05320}\ } (\bibinfo {year} {2015})}\BibitemShut {NoStop}%
\bibitem [{\citenamefont {Stefani}\ and\ \citenamefont
  {Kirillov}(2015)}]{stefani2015destabilization}%
  \BibitemOpen
  \bibfield  {author} {\bibinfo {author} {\bibfnamefont {F.}~\bibnamefont
  {Stefani}}\ and\ \bibinfo {author} {\bibfnamefont {O.~N.}\ \bibnamefont
  {Kirillov}},\ }\href@noop {} {\bibfield  {journal} {\bibinfo  {journal}
  {arXiv preprint arXiv:1506.00399}\ } (\bibinfo {year} {2015})}\BibitemShut
  {NoStop}%
\end{thebibliography}
\end{document}